\documentclass[conference]{IEEEtran}
\usepackage{amsmath}
\usepackage{latexsym}
\usepackage{graphicx}
\usepackage{bbding}
\usepackage{indentfirst}
\usepackage{cases}
\usepackage{algorithm2e}
\usepackage{subeqnarray}
\usepackage{slashbox}
\IEEEoverridecommandlockouts

\begin{document}

\title{Mode Selection in MU-MIMO Downlink Networks: A Physical Layer Security Perspective}
\author{\authorblockN{
Xiaoming~Chen, \emph{Senior Member, IEEE}, and Yu~Zhang,
\emph{Member, IEEE}
\thanks{Xiaoming~Chen ({\tt chenxiaoming@nuaa.edu.cn}) is with the
College of Electronic and Information Engineering, Nanjing
University of Aeronautics and Astronautics, Nanjing 210016, China. Yu~Zhang ({\tt
zhangyu\_wing@hotmail.com}) is with the Department of Information
Science and Electronic Engineering, Zhejiang University, Hangzhou
310027, China.}}}\maketitle

\begin{abstract}
In this paper, we consider a homogenous multi-antenna downlink
network where a passive eavesdropper intends to intercept the
communication between a base station (BS) and multiple secure users
(SU) over Rayleigh fading channels. In order to guarantee the
security of information transfer, physical layer security is
employed accordingly. For such a multiple user (MU) secure network,
the number of accessing SUs, namely transmission mode, has a great
impact on the secrecy performance. Specifically, on the one hand, a
large number of accessing SUs will arise high inter-user
interference at SUs, resulting in a reduction of the capacity of the
legitimate channel. On the other hand, high inter-user interference
will interfere with the eavesdropper and thus degrades the
performance of the eavesdropper channel. Generally speaking, the
harmful inter-user interference may be transformed as a useful tool
of anti-eavesdropping. The focus of this paper is on selecting the
optimal transmission mode according to channel conditions and system
parameters, so as to maximize the sum secrecy outage capacity.
Moreover, through asymptotic analysis, we present several simple
mode selection schemes in some extreme cases. Finally, simulation
results validate the effectiveness of the proposed mode selection
schemes in MU secure communications.
\end{abstract}


\section{Introduction}
Information security is always a critical issue of wireless
communications due to the open nature of wireless channels.
Traditionally, information security is realized by using complex
cryptography technology. In fact, information theory has proven that
secure communication can be guaranteed by only exploiting the
characteristics of wireless channels, e.g., fading, noise and
interference, namely physical layer security \cite{Wyner}
\cite{PLS1}. The essential of physical layer security is to maximize
the secrecy rate, which is defined as the rate difference between
the legitimate channel and the eavesdropper channel
\cite{SecrecyCapacity1} \cite{SecrecyCapacity2}. If there are
multiple antennas at the information source, by exploiting spatial
degrees of freedom, the legitimate channel rate is increased and the
eavesdropper channel rate is decreased simultaneously, so the
secrecy rate can be improved significantly. Thus, physical layer
security coupling with multi-antenna techniques has received
considerably research interests
\cite{Multiantenna1}-\cite{Multiantenna3}.

\subsection{Related Works}
Intuitively, in multi-antenna secure networks, by transmitting the
information in the null space of the eavesdropper channel, the
eavesdropper can not intercept any useful information. However, in
the sensing of maximizing the secrecy rate, this approach seems not
to be optimal. The key is to select a transmit direction, namely
beamforming, so as to achieve an optimal tradeoff between maximizing
the legitimate channel capacity and minimizing the eavesdropper
channel capacity \cite{Beamforming1}-\cite{Beamforming3}. In
\cite{Multiantenna4}, the problem of optimal transmit beamforming in
a MISO system was addressed by maximizing the secrecy rate. A
potential drawback of the above approach lies in that the source
must have full channel state information (CSI) to design the
transmit beam. To alleviate the assumption, a joint power allocation
and beamforming scheme was proposed based on full CSI of the
legitimate channel and partial CSI of the eavesdropper channel
\cite{Multiantenna5}. Yet, it is difficult to obtain the CSI for the
source, especially the CSI of the eavesdropper channel, because the
passive eavesdropper will hide itself as good as possible. It is
proved that if there is no CSI of the eavesdropper channel, the
beamforming alone the direction of the legitimate channel is optimal
\cite{LimitedFeedback2}. Since the secrecy rate is jointly
determined by the legitimate and the eavesdropper channel
capacities, if the source has no CSI of the eavesdropper channel, it
is impossible to maintain a steady secrecy rate over all
realizations of fading channels. In this context, the secrecy outage
capacity is adopted as a useful and intuitive metric to evaluate
security, which is defined as the achievable maximum secrecy rate
under the condition that the outage probability that the real
transmission rate surpasses the secrecy rate is equal to a given
value \cite{Outagecapacity1}. The secrecy outage capacity based on
antenna selection is analyzed in an uncorrelated MIMO system
\cite{Outagecapacity2} and in a correlated MIMO system
\cite{Outagecapacity3}. Note that the assumption of full CSI for the
legitimate channel at the source also seems impractical in
multi-antenna systems, especially in frequency division duplex (FDD)
systems. To solve this problem, limited feedback techniques are
introduced into multi-antenna secure systems to convey the quantized
CSI from the legitimate receiver to the source
\cite{LimitedFeedback1} \cite{LimitedFeedback2}.

Another advantage of the MIMO system lies in that it can support
multiple users transmission, namely space division multiple access,
so the performance is improved significantly
\cite{SDMA1}-\cite{SDMA3}. In \cite{MultiuserPLS1}, the secrecy rate
over MU-MIMO broadcasting channels was well studied. In
\cite{MultiuserPLS2}, a robust beamforming scheme for MU-MIMO
downlink networks was proposed by using a Bayesian approach. In
MU-MIMO systems, inter-user interference is a pivotal factor
affecting the overall performance. For a given precoding scheme, a
large number of accessing users is beneficial to exploit the spatial
multiplexing gain, but also arises high inter-user interference at
users. Thus, it is better to select the number of accessing users,
namely transmission mode, according to channel conditions and system
parameters \cite{Modeselection1} \cite{Modeselection2}.
Interestingly, in MU secure communications, the harmful inter-user
interference can be used to interfere with the interception of the
eavesdropper. Generally speaking, inter-user interference has two
completely opposite functions. A detailed investigation of the
impact of multiuser interference on the secrecy performance was
carried out in \cite{MultiuserInterference}, and thus multiuser
scheduling was called for secrecy performance enhancement. A single
user selection scheduling was proposed in
\cite{Multiuserscheduling}, so the inter-user interference can be
avoided completely. On the contrary, the users that the
multi-antenna system can support at best was scheduled in
\cite{HomoNet}, so as to improve the sum rate of the legitimate
channel. In fact, the number of scheduled users should be carefully
selected according to channel conditions. Therefore, it makes sense
to perform mode selection in MU-MIMO downlink networks from a
perspective of optimizing the performance.

\subsection{Main Contributions}
In this paper, we consider a MU-MIMO downlink network in presence of
a passive eavesdropper. Considering the large CSI feedback amount
for multiple legitimate channels, opportunistic space division
multiple access (OSDMA) \cite{OSDMA1} \cite{OSDMA2} is adopted to
exploit the MU gain due to its low complexity, small overhead and
good performance. The focus of this paper is on mode selection to
optimize the utility of inter-user interference in MU secure
communications based on physical layer security. The major
contributions of this paper can be summarized as follows:
\begin{enumerate}
\item We present a framework of physical layer security in MU-MIMO
downlink networks with limited CSI feedback based on OSDMA, and
propose to transform the harmful inter-user interference to enhance
wireless security through user scheduling.

\item It is found that under different channel conditions, inter-user
interference plays different roles. We derive an explicit expression
for secrecy outage capacity in terms of transmission mode, transmit
power, user number, and channel condition. By maximizing the sum
secrecy outage capacity, we obtain an adaptive mode selection
scheme.

\item Through asymptotic analysis, we provide some
guidelines for simple mode selection as follows:

\begin{enumerate}

\item At low SNR regime, maximum available
mode should be adopted. Relaxing the requirement on the outage
probability and increasing the number of SUs are hardly helpful to
improve the secrecy performance.

\item At high SNR regime, single SU mode is the best
choice and multiple-SU mode will result in zero rate.

\item If the number of SUs is sufficiently large, maximum available
mode can asymptotically achieve the optimal secrecy performance.

\end{enumerate}

\end{enumerate}

\subsection{Paper Organization}
The rest of this paper is organized as follows: Section II gives a
brief introduction of the considered MU-MIMO downlink network
employing physical layer security. Section III focuses on the
analysis and the design of the mode selection scheme based on the
criterion of maximizing the sum secrecy outage capacity. Through
asymptotic performance analysis, we present several simple mode
selection schemes in some extreme cases in Section IV. Section V
presents several simulation results to validate the effectiveness of
the proposed schemes, and finally Section VI concludes the whole
paper.

\emph{Notations}: We use bold upper (lower) letters to denote
matrices (column vectors), $(\cdot)^H$ to denote conjugate
transpose, $E[\cdot]$ to denote expectation, $\|\cdot\|$ to denote
the $L_2$ norm of a vector, $|\cdot|$ to denote the absolute value,
$(a)^{+}$ to denote $\max(a,0)$, $\lceil a\rceil$ to denote the
smallest integer not less than $a$, $\lfloor a\rfloor$ to denote the
largest integer not greater than $a$, and $\stackrel{d}{=}$ to
denote the equality in distribution. The acronym i.i.d. means
``independent and identically distributed", pdf means ``probability
density function" and cdf means ``cumulative distribution function".

\section{System Model}
\begin{figure}[h] \centering
\includegraphics [width=0.4\textwidth] {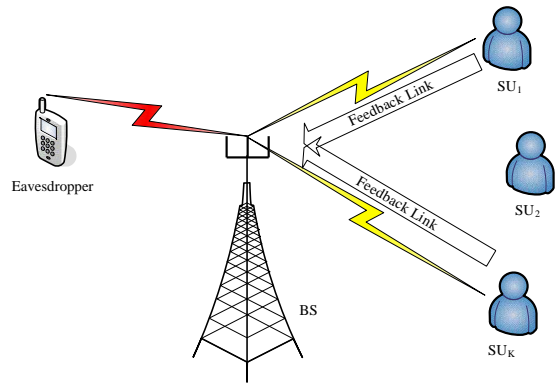}
\caption {A model of the MU-MIMO downlink network with physical
layer security.} \label{Fig1}
\end{figure}

We consider a MU-MIMO downlink network, where a base station (BS)
with $N_t$ antennas communicates with $K$ single antenna secure
users (SU), while a single antenna eavesdropper also receives the
message sent from the BS and tries to detect it. We use
$\textbf{h}_k$ to denote the $k$th legitimate channel from the BS to
the $k$th SU, whose elements are i.i.d. zero mean and unit variance
complex Gaussian random variables (homogeneous network in terms of
users's average channels as in \cite{HomoNet}). Similarly, we use
$\alpha\textbf{g}$ to denote the eavesdropper channel from the BS to
the eavesdropper, where $\alpha$ is the relative path loss and
$\textbf{g}$ denotes the small-scale fading vector with i.i.d. zero
mean and unit variance complex Gaussian distributed entries. We
assume that the BS has partial instantaneous CSI about the
legitimated channel through limited feedback and only statistical
CSI about the eavesdropper channel, since the eavesdropper is
passive.

The network is operated in the form of time slots. It is assumed
that the channels remain constant during a time slot, and
independently fade slot by slot. At the beginning of each time slot,
the BS randomly generates $M$ $N_t$-dimensional normalized
orthogonal vectors $\textbf{w}_{m},m=1,\cdots, M$, where $M$ with
the constraint of $1\leq M\leq N_t$ is the so-called transmission
mode. For example, select $M$ vectors out of the $N_t$ singular
vectors of a $N_t\times N_t$ complex Gaussian random matrix. The
selection of $M$, namely mode selection, is the focus of this paper,
and we will discuss it later in detail. Then, the BS broadcasts a
precoded training vector
$\textbf{x}=\sum\limits_{m=1}^{M}\textbf{w}_ms_i$ to the SUs, where
$s_i$ is the normalized training symbol known by the SUs in advance.
It is assumed that each user has perfect CSI related to its
corresponding legitimate channel through channel estimation and the
information feedback is error-free. Then, the $k$-th SU computes its
received signal-to-interference-plus-noise ratio (SINR) over the
$m$-th beam as
\begin{equation}
\gamma_{k,m}=\frac{P|\textbf{h}_{k}\textbf{w}_m|^2}{\sum\limits_{i=1,i\neq
m}^{M}P|\textbf{h}_{k}\textbf{w}_i|^2+\sigma^2},\label{eqn1}
\end{equation}
where $P$ is the transmit power on each beam and $\sigma^2$ is the
noise variance. By comparing its $M$ SINRs over $M$ beams, the
$k$-th SU finds the optimal beam according to the following
criterion:
\begin{equation}
\textbf{w}_{m_k}=\arg\max\limits_{1\leq m\leq
M}\gamma_{k,m}.\label{eqn2}
\end{equation}
Then, the $k$-th SU conveys the index $m_k$ and the corresponding
SINR $\gamma_{k, m_k}$ to the BS. After receiving the feedback
information from the $K$ SUs, the BS selects an optimal SU with the
largest SINR for each beam. Thereafter, the BS communicates with the
selected $M$ SUs in the rest of this time slot. It is worth pointing
out the probability of a SU being selected by multiple beams if the
number of SU is large, so we neglect it in this paper. For
convenience, we use $\textbf{h}_m$ to denote the SU's channel vector
over the $m$-th beam $\textbf{w}_m$, then the corresponding
legitimate channel capacity and the eavesdropper channel capacity
are given by
\begin{eqnarray}
C_{l,m}=\log_2(1+\lambda_m),\label{eqn3}
\end{eqnarray}
and
\begin{equation}
C_{e,m}=\log_2(1+\eta_m),\label{eqn4}
\end{equation}
respectively, where
$\lambda_m=\frac{|\textbf{h}_{m}\textbf{w}_m|^2}{\sum\limits_{i=1,i\neq
m}^{M}|\textbf{h}_{m}\textbf{w}_i|^2+\sigma^2/P}$ and
$\eta_m=\frac{|\textbf{g}\textbf{w}_m|^2}{\sum\limits_{i=1,i\neq
m}^{M}|\textbf{g}\textbf{w}_i|^2+\sigma^2/\alpha^2P}$. Thus, the
secrecy rate on the $m$-th beam can be expressed as \cite{HomoNet}
\begin{equation}
C_{sec,m}=\left(C_{l,m}-C_{e,m}\right)^{+}.\label{eqn5}
\end{equation}

Since the BS have no knowledge of the eavesdropper channel
$\textbf{g}$, it is impossible to maintain a steady secrecy rate
over all realizations of the fading channel. In this case, the BS
can only transmit the signal with a fixed rate. Thus, there
inevitably exists the case that the transmission rate is greater
than the secrecy rate. Then, the information may be intercepted by
the eavesdropper. In order to guarantee the security of information
transmission, the probability that the transmission rate surpasses
the secrecy rate, namely the outage probability, must be controlled
within a bearable range. In previous literatures, the achievable
maximum secrecy rate fulfilling a given requirement on the outage
probability $\varepsilon$ is called as secrecy outage capacity
$R_m(\varepsilon)$. Mathematically, it is given by
\begin{equation}
P_r(R_m(\varepsilon)>C_{sec,m})=\varepsilon.\label{eqn6}
\end{equation}

In this paper, we take the sum secrecy outage capacity as the
performance metric. As the name implies, sum secrecy outage capacity
denotes the sum of all scheduled SUs' secrecy outage capacity. Note
that given a requirement on the outage probability $\varepsilon$, an
important factor affecting the sum secrecy outage capacity is the
number of accessing users $M$, namely transmission mode. A large $M$
means higher inter-user interference, resulting in a decrease of
both the legitimate and the eavesdropper channel capacities.
Therefore, there exists an optimal $M$ maximizing the sum secrecy
outage capacity. In what follows, we focus on the optimal mode
selection for such a MU-MIMO downlink network, so as to maximize the
sum secrecy outage capacity.

\section{Optimal Mode Selection}
In this section, we concentrate on dynamic mode selection in a
MU-MIMO downlink network according to channel conditions and
security requirements, so as to maximize the sum secrecy outage
capacity. Prior to discussing the mode selection, we first give a
quantitative analysis of the secrecy outage capacity to reveal the
effect of transmission mode. From (\ref{eqn6}), the outage
probability on the $m$-th beam can be calculated as
\begin{eqnarray}
\varepsilon&=&P_r\left(R_m(\varepsilon)>\log_2\left(\frac{1+\lambda_m}{1+\eta_m}\right)\right)\nonumber\\
&=&P_r\left(\lambda_m<(1+\eta_m)2^{R_m(\varepsilon)}-1\right)\nonumber\\
&=&\int_0^{\infty}\int_0^{(1+\eta_m)2^{R_m(\varepsilon)}-1}f_{\lambda_m}(x)f_{\eta_m}(y)dxdy\nonumber\\
&=&\int_0^{\infty}F_{\lambda_m}\left((1+y)2^{R_m(\varepsilon)}-1\right)f_{\eta_m}(y)dy,\label{eqn7}
\end{eqnarray}
where $f_{\lambda_m}(x)$ and $F_{\lambda_m}(x)$ are the probability
density function (pdf) and cumulative distribution function (cdf) of
$\lambda_m$, respectively, and $f_{\eta_m}(y)$ is the pdf of
$\eta_m$. Clearly, in order to derive the outage probability, the
key is to obtain the distributions of $\lambda_m$ and $\eta_m$. In
the following, we turn our attention to the analysis of the two
distributions.

According to the definition of $\lambda_m$, it can be considered as
the maximum SINR on the $m$-th beam through choosing the optimal
channel $\textbf{h}_m$ out of $K$ ones. For an arbitrary channel
vector $\textbf{h}_n$ with i.i.d. zero mean and unit variance
complex Gaussian distributed entries, $|\textbf{h}_n\textbf{w}_m|^2$
is $\chi^2(2)$ distributed \cite{Distribution}, and thus
$\sum\limits_{i=1,i\neq m}^{M}|\textbf{h}_n\textbf{w}_i|^2$ is
$\chi^2(2M-2)$ distributed. Therefore, for a random variable
$\xi=\frac{|\textbf{h}_{n}\textbf{w}_m|^2}{\sum\limits_{i=1,i\neq
m}^{M}|\textbf{h}_{n}\textbf{w}_i|^2+\sigma^2/P}$, its cdf
$F_{\xi}(x)$ can be derived as
\begin{eqnarray}
F_{\xi}(x)&=&\int_0^{\infty}\int_0^{x(y+\sigma^2/P)}\frac{\exp(-y)y^{M-2}}{\Gamma(M-1)}\exp(-z)dzdy\nonumber\\
&=&1-\frac{\exp(-x/\rho)}{(1+x)^{M-1}},\label{eqn8}
\end{eqnarray}
where $\rho=P/\sigma^2$ is the transmit SNR. Since $\lambda_m$ is
the maximum value from $K$ independent random variables distributed
as $\xi$, we have
\begin{eqnarray}
F_{\lambda_m}(x)&=&(F_{\xi}(x))^{K}\nonumber\\
&=&\left(1-\frac{\exp(-x/\rho)}{(1+x)^{M-1}}\right)^{K}.\label{eqn9}
\end{eqnarray}
For $\eta_m$, since the selection of $\textbf{w}_m$ is independent
of the eavesdropper channel $\textbf{g}$, so it has the cdf similar
to $\xi$, and thus its pdf can be expressed as
\begin{eqnarray}
f_{\eta_m}(y)&=&\frac{\partial\left(1-\frac{\exp(-y/\alpha^2\rho)}{(1+y)^{M-1}}\right)}{\partial y}\nonumber\\
&=&\frac{(M-1)\exp(-y/\alpha^2\rho)}{(1+y)^M}+\frac{\exp(-y/\alpha^2\rho)}{\alpha^2\rho(1+y)^{M-1}}.\label{eqn10}
\end{eqnarray}
Substituting (\ref{eqn9}) and (\ref{eqn10}) into (\ref{eqn7}), the
outage probability is transformed as (\ref{eqn11}) at the top of
next page,
\begin{figure*}
\begin{eqnarray}
\varepsilon&=&\int_{0}^{\infty}\left(1-\frac{\exp(-(2^{R_m(\varepsilon)}-1)/\rho)\exp(-(2^{R_m(\varepsilon)}/\rho)y)}{2^{(M-1)R_m(\varepsilon)}(1+y)^{M-1}}\right)^{K}\frac{(M-1)\exp(-y/\alpha^2\rho)}{(1+y)^M}dy\nonumber\\
&&+\int_{0}^{\infty}\left(1-\frac{\exp(-(2^{R_m(\varepsilon)}-1)/\rho)\exp(-(2^{R_m(\varepsilon)}/\rho)y)}{2^{(M-1)R_m(\varepsilon)}(1+y)^{M-1}}\right)^{K}\frac{\exp(-y/\alpha^2\rho)}{\alpha^2\rho(1+y)^{M-1}}dy\nonumber\\
&=&1+\int_0^{\infty}\sum\limits_{n=1}^{K}{K\choose
n}(-1)^n(M-1)a(n)\frac{\exp\left(-\left(\left(n2^{R_m(\varepsilon)}+1/\alpha^2\right)/\rho\right)y\right)}{(1+y)^{n(M-1)+M}}dy\nonumber\\
&&+\int_0^{\infty}\sum\limits_{n=1}^{K}{K\choose
n}(-1)^n\frac{a(n)}{\alpha^2\rho}\frac{\exp\left(-\left(\left(n2^{R_m(\varepsilon)}+1/\alpha^2\right)/\rho\right)y\right)}{(1+y)^{(n+1)(M-1)}}dy\nonumber\\
&=&1+\sum\limits_{n=1}^{K}{K\choose
n}(-1)^n\bigg((M-1)a(n)W\left(\mu(n),\nu(n)\right)
+\frac{a(n)}{\alpha^2\rho}W(\mu(n),\upsilon(n))\bigg).\label{eqn11}
\end{eqnarray}
\end{figure*}
where
$a(n)=\frac{\exp(-n(2^{R_m(\varepsilon)}-1)/\rho)}{2^{n(M-1)R_m(\varepsilon)}}$,
$\mu(n)=(n2^{R_m(\varepsilon)}+1/\alpha^2)/\rho$, $\nu(n)=n(M-1)+M$
and $\upsilon(n)=(n+1)(M-1)$. (\ref{eqn11}) is derived according to
[27, Eqn. 3.3532]. $W(x,N)$ with $N$ being a natural number, is
defined as
\begin{displaymath}
W(x,N)=\left\{\begin{array}{ll} 1/x & \textrm{if
$N=0$}\\
-\exp(x)\textmd{E}_{\textmd{i}}(-x) & \textrm{if $N=1$},\\
\frac{1}{\Gamma(N)}\sum\limits_{n=1}^{N-1}\Gamma(n)(-x)^{N-1-n}\\\quad-\frac{(-x)^{N-1}}{\Gamma(N)}\exp(x)\textmd{E}_{\textmd{i}}(-x)
& \textrm{if $N\geq2$}\end{array}\right.
\end{displaymath}
where
$\textmd{E}_{\textmd{i}}(x)=\int_{-\infty}^{x}\frac{\exp(t)}{t}dt$
is the exponential integral function. Since $\varepsilon$ is a
monotonously increasing function of $R_m(\varepsilon)$, for a given
$\varepsilon$, it is able to find the associated $R_m(\varepsilon)$
with a certain transmission mode $M$ according to (\ref{eqn11}). For
convenience, we use $G(R_m(\varepsilon))$ to represent
(\ref{eqn11}), and thus $G^{-1}(\varepsilon)$ is equivalent to
$R_m(\varepsilon)$, where $G^{-1}(x)$ indicates the inversive
function of $G(x)$. From (\ref{eqn11}), we can also derive the
interception probability that the secrecy rate is less than zero by
letting $R_m(\varepsilon)=0$. Mathematically, it can be expressed as
(\ref{eqn12}) at the top of next page.
\begin{figure*}
\begin{eqnarray}
P_r(C_{sec,m}<0)&=&G(0)\nonumber\\
&=&1-\sum\limits_{n=1}^{K}{K\choose
n}(-1)^n\bigg((M-1)W\left((n+1/\alpha^2)/\rho,\nu(n)\right)
\frac{1}{\alpha^2\rho}W((n+1/\alpha^2)/\rho,\upsilon(n))\bigg).\label{eqn12}
\end{eqnarray}
\end{figure*}

For such a homogeneous network, if the SUs have a common requirement
on the outage probability $\varepsilon$, the sum secrecy outage
capacity with transmission mode $M$ is given by
\begin{equation}
R=MG^{-1}(\varepsilon).\label{eqn13}
\end{equation}

In this paper, we intend to select an optimal transmission mode
$M^{\star}$, so as to maximize the sum secrecy outage capacity.
However, due to the complexity of (\ref{eqn11}), it is difficult to
present an explicit expression for $M^{\star}$. As we know, given
the number of BS antennas $N_t$, the transmission mode $M$ belongs
to $[1,N_t]$. In practical systems, the number of BS antennas is
quite limited, i.e., $N_t=4$ in LTE systems and $N_t=8$ in LTE-A
systems. Thus, we could first derive the secrecy outage capacity for
each mode, then determine the optimal mode with the maximum sum
secrecy outage capacity. The whole procedure can be summarized as
below.
\begin{algorithm}[Mode Selection Algorithm]
\caption{Mode Selection Algorithm} \KwIn{$N_t$, $P$, $\sigma^2$,
$\alpha^2$, $K$, and $\varepsilon$. $m=1$ and $\Delta{R}$ is a small
positive real value.} \KwOut{$M^{\star}$}
\While{$m\leq N_t$}{$R_m=0$;\\
\While{$G(R_m)<\varepsilon$}{$R_m=R_m+\Delta{R}$;} $m=m+1$;}
$M^{\star}=\arg\max\limits_{1\leq m\leq N_t}(mR_m)$.
\label{alg:mine}
\end{algorithm}

\emph{Remark}: Through mode selection, we find the optimal tradeoff
among spatial multiplexing gain, inter-user interference and
anti-eavesdropping, so the sum secrecy outage capacity is maximized.
As a simple example, if the SNR is sufficiently high, multiuser
transmission may lead to performance saturation due to inter-user
interference, so the secrecy performance is impossible to be
improved by increasing the SNR. In this context, single user
transmission, namely $M=1$, may be optimal in the sense of
maximizing the sum secrecy outage capacity. Moreover, it is worth
pointing out that Algorithm 1 can be extended to a general case with
an arbitrary detection technique. Specifically, for a certain
detection technique, such as successive interference cancellation,
we can derive the corresponding sum secrecy outage capacity or the
other secrecy performance metrics, which is always a function of the
transmission mode. Similarly, by optimizing the secrecy performance,
it is possible to obtain the optimal transmission mode.

\section{Asymptotic Analysis}
In order to reduce the complexity of mode selection, we perform
asymptotical analysis in some extreme cases, such as noise-limited
case, interference-limited case and large SU number case. In what
follows, we investigate the three cases, respectively.

\subsection{Noise-Limited Case}
If transmit power $P$ is quite small or the noise variance
$\sigma^2$ is large enough, the interference terms of $\lambda_m$
and $\eta_m$ can be negligible with respect to the noise term,
namely the noise-limited case. In this case, we obtain a simple mode
selection scheme as follows:

\emph{Theorem 1}: For the noise-limited case, full spatial
multiplexing, namely $M^{\star}=N_t$, can obtain the maximum sum
secrecy outage capacity.

\begin{proof}
Please refer to Appendix I.
\end{proof}

\emph{Remark}: From the Theorem 1, it is known that at extreme low
SNR regime, the sum secrecy outage capacity with different numbers
of SUs and/or different requirements on the outage probabilities
asymptotically approaches zero as the SNR decreases. This is because
as the SNR tends to zero, both the legitimate and eavesdropper
channel rates approaches to zero, then the secrecy rate becomes zero
regardless of the number of SUs and the requirement on the outage
probability.

Moreover, the interception probability that the secrecy rate is less
than zero in such a case can be expressed as (\ref{eqn17}).
\begin{figure*}
\begin{eqnarray}
P_r(C_{sec,m}<0)&=&\frac{\exp\left(\frac{1+1/\alpha^2}{\rho}\right)\left(1-\left(1-\exp(-(1+1/\alpha^2)/\rho)\right)^{K+1}\right)}{K+1}.\label{eqn17}
\end{eqnarray}
\end{figure*}
As $K$ approaches infinity, the interception probability in
(\ref{eqn17}) approaches zero, so the probability of nonzero secrecy
rate is nearly equal to 1. Then, as long as $K$ is large enough,
there is nonzero secrecy rate with probability 1.

\subsection{Interference-Limited Case}
If the transmit power $P$ is quite large or the noise variance
$\sigma^2$ is sufficiently small, the noise term is negligible with
respect to the interference term in the received SINRs $\lambda_m$
and $\eta_m$, namely the interference-limited case. In this case, we
also have a simple mode selection scheme as follows:

\emph{Theorem 2}: For the interference-limited case, single SU
transmission mode, namely $M^{\star}=1$, can obtain the maximum sum
secrecy outage capacity.

\begin{proof}
Please refer to Appendix II.
\end{proof}

\emph{Remark}: From the Theorem 2, it is known that at extreme high
SNR regime, the sum secrecy outage capacity with $M>1$
asymptotically approaches zero as the SNR increases. This is because
at high SNR, both the legitimate and eavesdropper channel rates with
$M>1$ have the same performance ceiling due to interference
limitation. In this context, the sum secrecy outage capacity tends
to zero.

In such a case, the interception probability is given by
\begin{equation}
P_r(C_{sec,m}<0)=\frac{1}{K+1}.\label{eqn22}
\end{equation}
More interestingly, it is found that the interception probability
$P_r(C_{sec,m}<0)$ is independent of the channel condition, and is a
monotonously decrease function of $K$. Then, it is possible to
enhance wireless security by adding the SUs.

\subsection{Large SU number case}
When the number of SU $K$ is large enough, one can always find $M$
SUs with orthogonal channels, such that the inter-user interference
is canceled. In this case, we present a simple mode selection scheme
as follows:

\emph{Theorem 3}: For the large SU number case, $M^{\star}=N_t$ can
obtain the maximum sum secrecy outage capacity.

\begin{proof}
Please refer to Appendix III.
\end{proof}

This large $K$ case cancels the interference completely, which is
equivalent to the noise-limited case, so they have the same optimal
transmission mode. Furthermore, the interception probability in this
case can be derived as
\begin{eqnarray}
P_r(C_{sec,m}<0)&=&2^{-(N_t-1)\log_2(1+\rho\ln(KN_t))}\nonumber\\
&\times&\exp\left(-\frac{2^{\log_2(1+\rho\ln(KN_t))}-1}{\alpha^2\rho}\right).\label{eqn24}
\end{eqnarray}
The interception probability decreases as $K$ and $N_t$ increase and
$\alpha^2$ decreases.

\section{simulation Results}
To evaluate the performance of the proposed transmission mode
selection scheme for a MU-MIMO downlink network employing with
physical layer security, we present several simulation results in
different scenarios. For convenience, we set $N_t=4$,
$\alpha^2=0.01$, $K=10$, $\varepsilon=0.05$ and $\Delta R=0.01$ for
all simulation scenarios without explicit explanation. In the
following, we use AMS to denote the proposed adaptive mode selection
scheme, and use FTM1 and FTM2 to denote the traditional fixed
transmission mode schemes with $M=1$ and $M=N_t$, respectively. In
addition, we use TSNR in dB to represent the transmit SNR
$10\log_{10}\frac{P}{\sigma^2}$. Without loss of generality, we take
the sum secrecy outage capacity as the performance metric.

\begin{figure}[h] \centering
\includegraphics [width=0.5\textwidth] {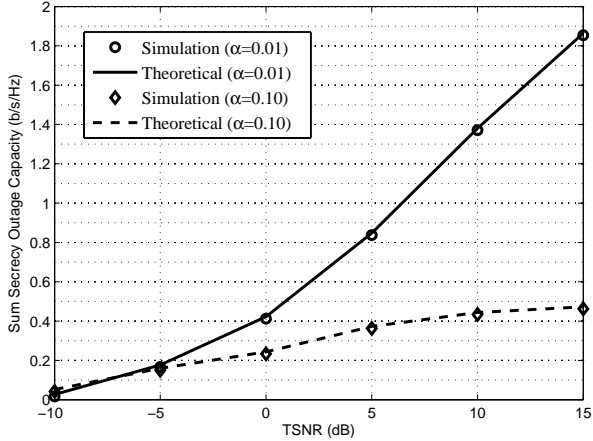}
\caption {Theoretical and simulation performance comparison with
different path losses.} \label{Fig5}
\end{figure}

First, we check the accuracy of the derived theoretical results with
different path losses. As shown in Fig.\ref{Fig5}, the theoretical
results are well coincided with the simulation results in the whole
TSNR region. It is found that at low TSNR, the sum secrecy outage
capacities with different path losses are nearly the same. This is
because the sum secrecy outage capacity asymptotically tends to zero
under this condition. However, as TSNR increases, the sum secrecy
outage capacity with $\alpha=0.01$ is obviously better than that
with $\alpha=0.10$, since the interception capability of the
eavesdropper becomes weak. So far, short-distance interception is
still an open issue for physical layer security.

\newcommand{\tabincell}[2]{\begin{tabular}{@{}#1@{}}#2\end{tabular}}
\begin{table*}\centering
\caption{Transmission modes for different schemes.} \label{Tab1}
\begin{tabular*}{9.48cm}{|c|c|c|c|c|c|c|c|c|c|c|c|}\hline
\backslashbox{Scheme}{TSNR}& -10 & -8 & -6 & -4 & -2 & 0 & 2 & 4 & 6 & 8 & 10\\
\hline AMS &  4  & 4 & 3 & 3 & 3 & 2 & 2 & 2 & 2& 1 & 1 \\
\hline FTM1 & 1 & 1 & 1 & 1 & 1 & 1 & 1 & 1 & 1 & 1 & 1 \\
\hline FTM2 & 4 & 4 & 4 & 4 & 4 & 4 & 4 & 4 & 4 & 4 & 4 \\
\hline
\end{tabular*}
\end{table*}

\begin{figure}[h] \centering
\includegraphics [width=0.5\textwidth] {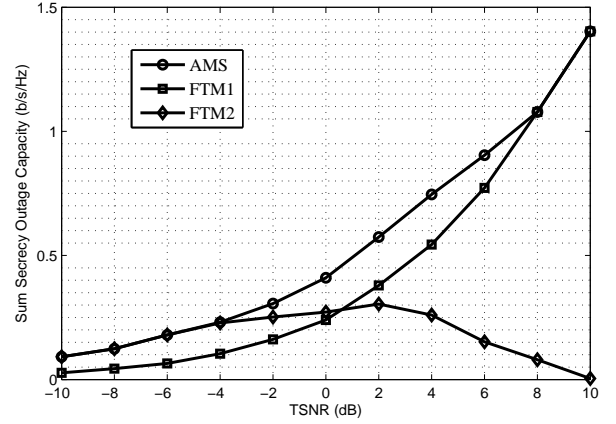}
\caption {Performance comparison of different mode selection
schemes.} \label{Fig2}
\end{figure}

Then, we compare the sum secrecy outage capacities of AMS, FTM1 and
FTM2 schemes. As seen in Tab.\ref{Tab1} at the top of next page, at
low SNR regime, the proposed AMS scheme chooses high order
transmission mode, since the inter-user interference is quite small
with respect to the noise in legitimate channel, and a large $M$ can
exploit the spatial multiplexing gain and counteract the
interception, so as to improve the sum secrecy outage capacity. For
example, as seen in Fig.{\ref{Fig2}}, there is about 0.2 b/s/Hz
performance gain over FTM1 with SNR $=-2$ dB. When SNR is greater
than a threshold, such as $8$dB, the proposed AMS scheme would adopt
single SU transmission mode, this is because it is
interference-limited under this condition, which is consistent with
our theoretical claim in Theorem 2. Additionally, at high SNR, the
sum secrecy outage capacity of FTM2 asymptotically approaches zero
as claimed in Proposition 2. Thus, the proposed AMS scheme can
achieve the optimal performance at the whole SNR regime, which is
helpful for MU-MIMO secure communications with physical layer
secrecy. In general, the proposed adaptive mode selection (AMS)
scheme determines the optimal transmission mode by comparing $N_t$
sum secrecy outage capacities, while FIM1 and FIM2 uses fixed
transmission modes regardless of channel conditions. Thus, AMS has
relative higher complexity than FIM1 and FIM2. However, since mode
selection is performed only when channel conditions change, not
within each time slot, the complexity of AMS is bearable in
practical systems. In addition, Theorem 1 and 2 can be used to
determine the transmission mode at low and high SNR regimes
respectively, which have the same complexity as FIM1 and FIM2.

\begin{table}\centering
\caption{Transmission modes for AMS with different outage
probabilities.} \label{Tab2}
\begin{tabular*}{8.83cm}{|c|c|c|c|c|c|c|c|c|c|c|c|}\hline
\backslashbox{$\varepsilon$}{TSNR}& -10 & -8 & -6 & -4 & -2 & 0 & 2 & 4 & 6 & 8 & 10\\
\hline 0.10 & 4 & 4 & 4 & 4 & 3 & 3 & 2 & 2 & 2 & 2 & 1 \\
\hline 0.05 & 4 & 4 & 3 & 3 & 3 & 2 & 2 & 2 & 2 & 1 & 1 \\
\hline 0.01 & 1 & 1 & 1 & 1 & 1 & 1 & 1 & 1 & 1 & 1 & 1 \\
\hline
\end{tabular*}
\end{table}

\begin{figure}[h] \centering
\includegraphics [width=0.5\textwidth] {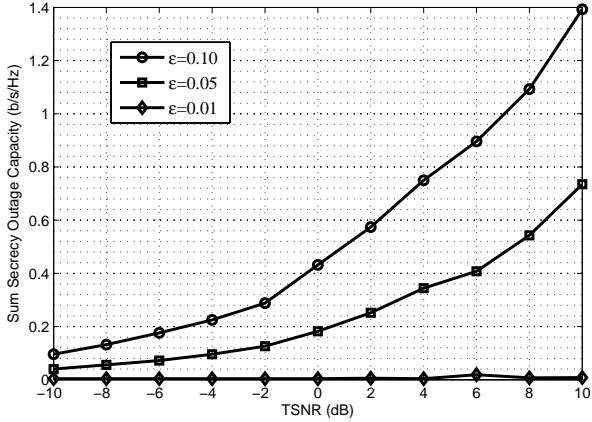}
\caption {Performance comparison of AMS scheme with different outage
probabilities.} \label{Fig3}
\end{figure}

Next, we investigate the impact of the outage probability on the sum
secrecy outage capacity of the proposed AMS scheme. The outage
probability $\varepsilon$ represents the interception probability
with a given secrecy rate. As shown in Fig.\ref{Fig3}, when
$\varepsilon=0.01$, namely a quite small interception probability,
the feasible secrecy outage capacity is nearly equal to zero. As
$\varepsilon$ increases, the sum secrecy outage capacity improves
accordingly. For example, one with $\varepsilon=0.10$ has about
$0.5$ b/s/Hz performance gain over the one with $\varepsilon=0.05$
at SNR $=6$ dB. Furthermore, the performance gain becomes larger
with the increase of SNR. Note that the sum secrecy outage
capacities with different outage probabilities approaches zero at
low SNR regime, which reconfirms the claims in Proposition 1. From
Tab.\ref{Tab2}, it is seen that as the requirement on the outage
probability relaxes, AMS will be apt to choose a high order
transmission mode, especially at low SNR regime.

\begin{table}\centering
\caption{Transmission modes for AMS with different numbers of SUs.}
\label{Tab3}
\begin{tabular*}{8.84cm}{|c|c|c|c|c|c|c|c|c|c|c|c|}\hline
\backslashbox{$K$}{TSNR}& -10 & -8 & -6 & -4 & -2 & 0 & 2 & 4 & 6 & 8 & 10\\
\hline 50 & 4 & 4 & 4 & 4 & 3 & 2 & 2 & 2 & 2 & 2 & 2 \\
\hline 20 & 4 & 4 & 4 & 4 & 3 & 2 & 2 & 2 & 2 & 2 & 1 \\
\hline 5  & 4 & 4 & 4 & 3 & 2 & 2 & 2 & 1 & 1 & 1 & 1 \\
\hline
\end{tabular*}
\end{table}

\begin{figure}[h] \centering
\includegraphics [width=0.5\textwidth] {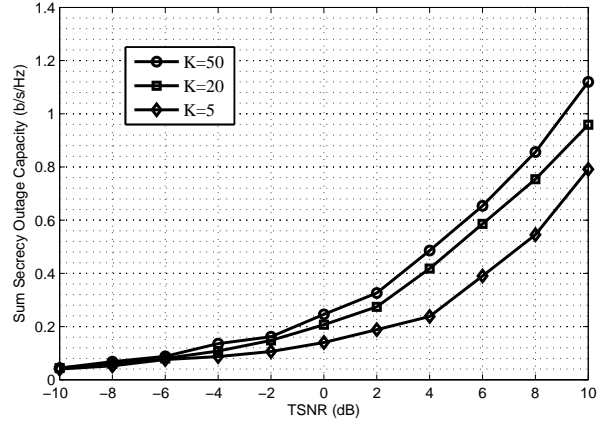}
\caption {Performance comparison of AMS scheme with different number
of SUs.} \label{Fig4}
\end{figure}

Finally, we show the benefit of the proposed AMS scheme from the
perspective of the SU number. As seen in Fig.\ref{Fig4}, with the
increase of the SU number $K$, the sum secrecy outage capacity
increases accordingly, since the probability that the channels of
the selected SUs are orthogonal becomes larger. In other words, the
inter-user interference in legitimate channel is smaller gradually
while the one in the eavesdropper channel remains unchaged in the
statistical sense. In addition, it is found that the performance
gain by increasing $K$ becomes smaller with a large $K$. For
example, at SNR = 8 dB, the performance gain by increasing 15 SUs
from $K=5$ to $K=20$ heavily surpasses that by increasing 30 SUs
from $K=20$ to $K=50$. It is because when $K$ is large, the received
SNR based on OSDMA approaches $\rho\ln(KN_tN_r)$ as analyzed in
Section IV.C, so the gain becomes smaller by increasing the same SUs
as $K$ increases. Moreover, the performance gain by adding SUs is
negligible at low SNR, which proves Proposition 1 again. Similarly,
from Tab.\ref{Tab3}, it is also seen that as the number of SUs
increases, AMS will be apt to choose a high order transmission mode,
especially at low SNR regime.

\section{Conclusion}
The main contribution of this paper is to exploit the benefit of
inter-user interference to improve the sum secrecy outage capacity
in MU-MIMO downlink networks employing physical layer security. It
is found that under different channel conditions, the inter-user
interference plays different roles. Therefore, we propose an
effectively adaptive transmission mode selection scheme maximizing
the sum secrecy outage capacity. Furthermore, asymptotic analysis is
carried out to further insights on mode selection. For example, our
asymptotical results show that at low SNR regime, maximum available
mode should be adopted. Both relaxing the requirement on the outage
probability and increasing the number of SUs are hardly helpful to
improve the secrecy performance. On the contrary, at high SNR
regime, single SU mode is the best choice and multiple-SU mode will
result in zero rate.

\begin{appendices}

\section{Proof of Theorem 1}
In the noise-limited case, the cdf of $\lambda_m$ and the pdf of
$\eta_m$ are reduced to
\begin{equation}
F_{\lambda_m}(x)=\left(1-\exp\left(-\frac{x}{\rho}\right)\right)^{K},\label{eqn14}
\end{equation}
and
\begin{equation}
f_{\eta_m}(y)=\frac{1}{\alpha^2\rho}\exp\left(-\frac{y}{\alpha^2\rho}\right),\label{eqn15}
\end{equation}
respectively. Substituting (\ref{eqn14}) and (\ref{eqn15}) into
(\ref{eqn7}), the outage probability in this case can be computed as
\begin{eqnarray}
\varepsilon&=&\int_{0}^{\infty}\left(1-\exp\left(-\frac{2^{R_m(\varepsilon)}-1}{\rho}\right)
\exp\left(-\frac{2^{R_m(\varepsilon)}}{\rho}y\right)\right)^{K}\nonumber\\
&&\times\frac{\exp\left(-\frac{y}{\alpha^2\rho}\right)}{\alpha^2\rho}
dy\nonumber\\
&=&1+\sum\limits_{n=1}^{K}{K\choose
n}(-1)^n\frac{\exp\left(-n\left(2^{R_m(\varepsilon)}+1/\alpha^2\right)/\rho\right)}{n2^{R_m(\varepsilon)}+1}.\label{eqn16}
\end{eqnarray}

Interestingly, it is found that the outage probability in
(\ref{eqn16}) is independent of the transmission mode. Under this
condition, $M=N_t$ can achieve the full multiplexing gain, and also
leads to the maximum sum secrecy outage capacity. Thus, we complete
the proof.

\section{Proof of Theorem 2}
In the interference-limited case, the cdf of $\lambda_m$ and the pdf
of $\eta_m$ can be expressed as
\begin{eqnarray}
F_{\lambda_m}(x)=\left(1-\frac{1}{(1+x)^{M-1}}\right)^{K},\label{eqn18}
\end{eqnarray}
and
\begin{eqnarray}
f_{\eta_m}(y)=\frac{M-1}{(1+y)^{M}}.\label{eqn19}
\end{eqnarray}
Similarly, the outage capacity is given by
\begin{eqnarray}
\varepsilon&=&\int_0^{\infty}\left(1-\frac{1}{2^{(M-1)R_m(\varepsilon)}(1+y)^{M-1}}\right)^{K}\frac{M-1}{(1+y)^{M}}dy\nonumber\\
&=&1+\sum\limits_{n=1}^{K}{K\choose
n}\frac{\left(-2^{-(M-1)R_m(\varepsilon)}\right)^n}{n+1}\nonumber\\
&=&\frac{2^{(M-1)R_m(\varepsilon)}\left(1-\left(1-2^{-(M-1)R_m(\varepsilon)}\right)^{K+1}\right)}{K+1}\label{eqn20}\\
&=&\frac{2^{\frac{M-1}{M}R}\left(1-\left(1-2^{-\frac{M-1}{M}R}\right)^{K+1}\right)}{K+1}.\label{eqn21}
\end{eqnarray}
where (\ref{eqn20}) is obtained according to [27, Eqn.0.1553]. From
(\ref{eqn21}), it is known that $\varepsilon$ is a monotonously
increasing function of $\frac{M-1}{M}$ and $R$. Given a requirement
on the outage probability $\varepsilon$, the sum secrecy outage
capacity $R$ is maximized by minimizing $\frac{M-1}{M}$. Clearly,
$\frac{M-1}{M}$ is minimized when $M=1$, which proves the claims in
Theorem 2.

\section{Proof of Theorem 3}
If the user number $K$ is large enough, there is the following
important property that $\lambda_m\rightarrow\rho\ln(KN_t)$
\cite{LargeK}. In other words, the legitimate channel capacity
approaches a constant $\log_2(1+\rho\ln(KN_t))$, so the outage
probability is transformed as
\begin{eqnarray}
\varepsilon&=&P_r\left(\eta_m>2^{\log_2(1+\rho\ln(KN_t))-R_m(\varepsilon)}-1\right)\nonumber\\
&=&\frac{\exp\left(-\frac{2^{\log_2(1+\rho\ln(KN_t))-R_m(\varepsilon)}-1}{\alpha^2\rho}\right)}{\left(2^{\log_2(1+\rho\ln(KN_t))-R_m(\varepsilon)}\right)^{M-1}}.\label{eqn23}
\end{eqnarray}
As seen in (\ref{eqn23}), $\varepsilon$ is a monotonously increasing
function of $R_m(\varepsilon)$ and is a monotonously decreasing
function of $M$. Given a requirement on the outage capacity
$\varepsilon$, secrecy outage capacity $R_m(\varepsilon)$ is
maximized with $M=N_t$, which also achieves the full multiplexing
gain. Thus, we get Theorem 3.

\end{appendices}

%
%

\end{document}